\documentstyle[12pt,aasms4]{article} \doublespace
\def\wig#1{\mathrel{\hbox{\hbox to 0pt{%
\lower.5ex\hbox{$\sim$}\hss}\raise.4ex\hbox{$#1$}}}}

\newcommand{\beq}{\begin{equation}} \newcommand{\eeq}{\end{equation}}
\newcommand{\eeql}[1]{\label{eq:#1}\end{equation}}

\newcounter{compteur}

\def\bib{\par\noindent\hangindent=3mm\hangafter=1}

\def\mv{M_V}
\def\mbol{M_{bol}}
 \def\lum{L/L_\odot} \def\msol{M_\odot}
   \def\te{T_{\rm
eff}}  \def\simgr{\,\hbox{\hbox{$ > $}\kern -0.8em
\lower 1.0ex\hbox{$\sim$}}\,} \def\simle{\,\hbox{\hbox{$ < $}\kern -0.8em
\lower 1.0ex\hbox{$\sim$}}\,} \def\wig#1{\mathrel{\hbox{\hbox to 0pt{%
\lower.5ex\hbox{$\sim$}\hss}\raise.4ex\hbox{$#1$}}}}

\begin{document}

\title{\bf Contribution of brown dwarfs and white dwarfs to recent
microlensing observations and to the halo mass
budget} \author{{\sc G. Chabrier, L. Segretain and D. M\'era}\\ Centre de
Recherche Astrophysique de Lyon (UMR CNRS 5574),\\ Ecole Normale
Sup\'erieure, 69364 Lyon Cedex 07, France\\}



\begin{abstract} We examine the recent results of the MACHO collaboration
towards the Large Magellanic Cloud (Alcock et al. 1996) in terms of a halo brown dwarf or
white dwarf
population. The possibility for most of the microlensing events to be due to brown
dwarfs is totally excluded by large-scale kinematic properties.
The white dwarf scenario
is examined in details in the context of
the most recent white dwarf cooling theory
(Segretain et al. 1994) which includes explicitely the extra
source of energy due to carbon-oxygen differentiation at
crystallization, and the subsequent Debye cooling.
We show that the observational constraints arising from the luminosity function of high-velocity white dwarfs
in the solar neighborhood and from the recent HST deep field counts are consistent with a white dwarf contribution
to the halo missing mass as large as 50\%, provided i) an IMF strongly peaked around $\sim 1.7\,\msol$ and ii) a halo
age older than 
$\sim$ 18 Gyr.

\bigskip \bigskip

Subject headings : stars: low-mass, brown dwarfs --- stars: white dwarfs --- stars : luminosity function, mass function --- The Galaxy : halo ---
dark matter

\end{abstract}


\section{Introduction} There is compelling evidence for believing that a large
amount, if not
most of baryonic matter in the Universe is under the form of dark,
unobserved objects.
On the other hand, there is also
evidence that spiral galaxies are surrounded by a large amount of
non-luminous mass which is responsible for their observed small-scale and
large-scale kinematic properties (velocity dispersion and rotation curve).
These two facts yield the natural conclusion that baryonic dark matter is
a plausible candidate for halo dark matter. A breakthrough in this
longstanding, unsolved problem has been accomplished recently with the
developement of microlensing experiments, by inferring the presence of
dark objects in the halo through their gravitational effect on luminous
matter. A detailed analysis of the first year of the MACHO (Alcock et
al. 1993) and EROS (Aubourg et al. 1993) observations towards the LMC,
complemented by the determination of the mass function of
low-mass halo stars, yielded an average mass for the dark objects
$<m>\approx 0.03\,\msol$, well within the brown dwarf domain (M\'era,
Chabrier and Schaeffer 1996a). The inferred maximum brown dwarf
contribution to the halo mass budget was found to be $\sim$ 10 to 20\%
(Alcock et al. 1995; Gates, Gyuk and Turner, 1995; M\'era et al. 1996a).
These results have now to be reconsidered in the light of the most recent
analysis of the MACHO collaboration, which includes one more year
LMC data. This yields now a total of {\it seven} candidates for MACHO
with {\it longer}
durations, from 30 to 110 days (Alcock et al. 1996).

In this Letter, we re-examine the M-dwarf,
brown dwarf and
white dwarf contributions to the halo mass budget
in light of the most recent observational constraints (star counts and MACHO results) and white dwarf cooling theory.

\section {M-dwarfs, brown dwarfs}

HST star counts at large magnitudes (Bahcall et al.
1994, Hu et al. 1994) show that the
M-dwarf contribution to the Galactic missing mass $M_{dyn}\approx$ $10^{12}\,\msol$ represents at most a few
percents. A more precise determination
can be obtained from the observed luminosity function (LF) of high-velocity
M-dwarfs in the solar neighborhood (Dahn et al. 1995). A detailed analysis
of this LF shows that the mass function (MF) of the
spheroid (characterized by a $\sim 1/r^3$
density-profile) is reasonably well described by a power-law MF
$\phi(m)\propto m^{-\alpha}$ with $\alpha \approx 2-2.5$, from 0.6 $\msol$
down to the hydrogen burning limit and a normalization
$dN/dm(0.1\,\msol)=10^{-2.7}\,\msol^{-1}$pc$^{-3}$ (M\'era et al. 1996a).
This yields a
{\it maximum}
local density for the spheroid+halo M-dwarf population $\sim 4\times 10^{-5}\,\msol$
pc$^{-3}$, and thus an optical depth $\tau_{Mdwarf}< 4\times 10^{-9}$. A comparison with the
value inferred from the new MACHO results $\tau \approx 1.7 \times 10^{-7}$
shows convincingly
that M-dwarfs can be responsible for less than $0.1$ of the
 microlensing events towards the LMC.
On the other hand, reasonnable estimates for the LMC characteristics yield an optical
depth $\tau \approx 5\times 10^{-8}$, a factor 3 to 4 smaller than the recent
MACHO results (Sahu 1994). 

As mentioned above, the analysis of the previous MACHO+EROS results showed
that the observed events were likely to be due to halo brown dwarfs, with
an average mass $<m>\approx 0.03\,\msol$ (M\'era et al. 1996a).
These calculations must be re-examined in the context of the
recent MACHO
results.
The new 7 events
yield $<t_e>\propto <\sqrt m><{1\over v}> \approx 40$ days (Alcock et al. 1996)\footnote{Note that the event time-scale is defined here as the
Einstein {\it radius} crossing time, whereas the MACHO group adopts the
Einstein {\it diameter} crossing time in their definitions.}.
Since the minimal tangential velocity $<v>$ of the
lens is bound by the rotation velocity of the line of sight ($\sim 220$ km s$^{-1}$)(M\'era et al. 1996b), this
excludes totally the possibility for these events
to be due to brown dwarfs and yields an average mass $<m>\approx
0.5\,\msol$.
Since M-dwarfs are
excluded,
the inferred optical depth $\tau\approx 1.7\times 10^{-7}$ means
that about 40\% of the hidden mass consists of {\it halo white dwarfs}.
\footnote{Neutron stars and stellar black holes must be rejected as a significant halo population
mainly on the basis of the
severe constraints arising from the observed metallicity and helium
abundances (Ryu,
Olive and Silk 1990). In any case
neutron stars and black holes are likely to contribute considerably less
than WDs to the dark mass fraction.}

\section{White dwarfs}

The white dwarf contribution to the halo missing mass has been examined
over the past by different authors. These studies focussed mainly on the
constraint arising from the Galaxy chemical evolution (Olive 1986; Ryu et al. 1990).
These authors show that WDs are unlikely candidates for providing the {\it entire} halo missing
mass. Recent calculation of
the expected radiation signature of the
progenitors in galactic halos at large redshift (Charlot and Silk 1995)
show that a white dwarf mass fraction larger than $\sim 10\%$
of the missing mass would be in conflict with observations. These results
will be commented below.

An independent, more stringent constraint, comes from the observed
white dwarf luminosity
function (WDLF) in the solar neighborhood, as considered
by Tamanaha, Silk, Wood and Winget (1990). However, although pointing the way,
these calculations 
 were based on simplified WD interior and a WD cooling theory aimed at describing the {\it disk}
WDLF, thus appropriate for objects younger, and thus warmer,
than the expected halo population. In
particular these calculations do not include a {\it complete} treatment of
crystallization (see below), which occurs around $\log
\lum \approx -3.5$ in WD interiors.
This affects substantially the cooling of
halo WDs
 and will modify significantly the expected halo WDLF.
\footnote{After the present calculations were completed, we were aware of similar
studies by Adams and Laughlin (1996). These calculations, however, are
based on crude (pure carbon) WD interior and the afore-mentioned cooling theory.}

 In this {\it Letter}, we use the most updated WD
cooling theory for
carbon/oxygen WDs, with the appropriate equation of state
both in the classical and in the quantum (crystal) regime (Segretain et
al., 1994; Chabrier, 1993) and a helium-rich atmosphere (Wood 1992), characteristic of most cool ($\te \wig < 6000$ K) so-called "DB" WDs.
As first suggested by Stevenson (1980), the gravitational
energy release due to carbon-oxygen {\it differentiation} at crystallization
affects drastically the subsequent cooling time of the star, thus changing
the luminosity for a given age (Segretain \& Chabrier 1993). A consistent
treatment of the crystallization phase diagram along WD evolution has been
derived recently by Segretain et al. (1994).
As shown by these authors, the
crystallization processes modify appreciably the WD cooling time
and then the WDLF for $\log \lum \wig < -4$, characteristic of old disk and halo WDs. The LF derived with this
theory yields an estimate for the age of the Galactic disk
$\tau_{disk}\approx 10.5-12$ Gyr, depending on the bolometric correction
used for the {\it observed} LF (Hernanz et al. 1994), about 20\%
larger than estimates based on cooling theories which do not include
the {\it complete} crystallization process (Wood 1992; see Segretain et al. 1994
\S4.1 for details).

The calculations proceed as in 
Hernanz et al. (1994). The WDLF reads :

$$n(L)=\int_{m_{inf}(L)}^{m_{sup}} \tau_{cool}(L,m)\times \psi
\bigl[t_h-t_{cool}(L,m)-t_{ms}(m)\bigr]\times \phi(m)dm\eqno(1)$$

Here $\tau_{cool}=dt_{cool}/dM_{bol}$ is the {\it characteristic} cooling
time, where $t_{cool}$ is the WD cooling time. $t_{ms}$ and $t_h$
denote respectively the age spent on the main sequence for the WD
progenitor and the age of the halo. The function $\phi (m)$ is the initial
mass function and $m_{inf}$ and $m_{sup}$ denote respectively the
minimum and the maximum mass of the WD progenitors which contribute at luminosity
$L$. Since the age of the halo is much larger than any time associated
with star
formation, the initial stellar formation rate $\psi(t)$ is well
approximated by a burst at $t=0$, i.e. a $\delta(t=0)$ function.
In that case eqn(1) reduces to :

$$n(L)={dt_{cool}\over dM_{bol}}\times \nu(t_h-t_{cool})\times {dm\over dt} \eqno(2)$$

where $\nu(t_h-t_{cool})$ represents the number of WDs formed at
$t=t_h-t_{cool}$, i.e. the number of stars with a main sequence lifetime
$t_{ms}=t_h-t_{cool}$. 
We verified that finite-time SFR, e.g. a constant SFR along $\Delta t\ne 0$ (a reasonnable representation of a continous series of burst
SFRs) or an exponential SFR yield very similar results. The progenitor-WD mass relation is $m_{WD}\approx 0.45\,+\,0.1\,m$ (Iben and Tutukov 1984).
The WDLF is normalized to :

$$\int n\, dM_{bol}=-2.5\int n\,\,{\rm d} \log(\lum)=X_{WD}\,\rho_{dyn}/<m_{WD}>
\,\,{\rm pc}^{-3}\eqno(3)$$

where $X_{WD}$ is the (sought) mass fraction under the form of WDs in the
halo of the Galaxy. As shown in eqn(2) the most
essential parameter in this calculation is the white dwarf cooling time
$t_{cool}$. We use the afore-mentioned WD cooling sequences calculated in Segretain et al.
(1994) and Garcia-Berro et al. (1996).

The second important parameter to be determined in Eq.(1) is the IMF
$\phi(m)$. As mentioned in \S 2, a severe constraint arises from the
recently determined mass-function (slope and normalization)
of halo M-dwarfs.
The predicted star counts obtained with this MF for a
spheroid($1/r^3$)+halo($1/r^2$) density profile are in perfect agreement with the
observations of the HST at large magnitude ($I\ge 25$) (M\'era et al.,
1996a). On the other hand, the observed halo
metallicity implies that stars above $m\wig >8\,\msol$, believed to be
type II Supernovae progenitors,
represent at most $\sim 1\%$ of the halo initial stellar population (Ryu et al. 1990).
These observational constraints show that,
 for WDs to contribute significantly to
the mass of the halo, the IMF must exhibit a strongly {\it bimodal}
behaviour and peak around some characteristic mass in the range
$\left[ m_{inf},\sim 8\,\msol\right]$.
The minimum mass corresponds to a main-sequence lifetime of the progenitor
equal to the age of the halo, i.e.
$m_{inf}\approx 0.9\,\msol$ for
$t=10$ to 25 Gyr.
Several functional forms can be advocated for such an (unknown) IMF.
We elected a simple cut-off
power-law function
$\phi(m)=dN/dm=A\,e^{ -(\bar {m}/ m )^{\beta_1}} m^{-\beta_2}$ (see e.g. Larson 1986).
This form mimics adequatly a strongly peaked IMF and is very similar to functional forms based on stellar formation
theory (Adams and Laughlin 1996). The IMF is normalized to :

$$\int_{m_{inf}}^{\sim 8\,\msol} \phi(m)m_{WD}(m)dm=X_{WD}\rho_{dyn},$$

which determines $A$ (for a given $X_{WD}$ and $m_{WD}(m)$ relation). The parameter-space for
($\bar {m},\beta_1$, $\beta_2$) is constrained by the required negligible number
of stars outside the
mass-range $\left[\sim 0.9\,\msol,\sim 8\,\msol\right]$ but different
values yield quantitatively different mass-distributions.
A large number of masses $\ge 2\,\msol$ would raise severe problems for the fraction of ejected gas and the subsequent
helium and metal galactic enrichment (Hegyi and Olive 1986; Ryu et al. 1990).
In order to examine the dependence
 of the IMF on the results, we thus considered two functions, namely
($\bar {m}=2.0,\beta_1=2.2$, $\beta_2$=5.15), peaked around
$\sim  1.3\,\msol$ (hereafter IMF1), and ($\bar {m}=2.7,\beta_1=2.2$, $\beta_2$=5.75), peaked around
$\sim  1.7\,\msol$ (hereafter IMF2).
The complete M-dwarf+WD IMF's fulfilling all the afore-mentioned
constraints are shown on Figure 1.

{\bf {\it Observational constraints}}. The LF of field WDs
has been obtained by
Liebert et al. (1988)
up to $M_V\approx 19$ (i.e. $\lum \wig > 10^{-5}-10^{-6}$, depending on the bolometric correction $BC_V$).
 The LF declines abruptdely for $\mv\approx 16$, which corresponds
to $\log \lum \approx -4.2$ to $-4.6$. As stated by these authors, {\it no}
WD was found at fainter magnitudes, with this or
with
other proper-motion samples, whereas stars up to $\mv= 19$, i.e.
three magnitudes fainter, have been observed with similar programs
(Liebert et al. 1988; Monet et al. 1992). Interestingly enough, five
WDs in the Liebert et al. sample have tangential velocities
$v_{tan}>250$ kms$^{-1}$ and $M_V\ge 13$ and thus are assignable to the halo sample (shown by filled circles on
Figure 2).

More recent constraints arise from the HST counts up to $I=26.3$ (Flynn et al. 1996).
For a WD mass
fraction
$X_{WD}\sim 10\%$, it is easy to show that the observed number of WDs in the HST
field implies that {\it halo} WDs must
have $\mv \approx \mbol \wig >14$.
This is consistent with the observed high-velocity WDs. The colors of the disk and halo WDs were taken to be
$0\wig < V-I \wig <2$, as suggested by the observations (Liebert et al. 1988;
vonHippel, Gilmore \& Jones 1995), and the bolometric correction $0\wig < BC_V \wig <1$ (Liebert et al. 1988; Bergeron, Saumon \& Wesemael 1995).

The observed WDLF is represented on Figure 2, with different halo WDLFs. The dotted line is the {\it disk} WDLF from Segretain et al.
(1994) for an age $\tau_d=10.5$ Gyr \footnote{based on a blackbody bolometric correction $BC_V$ for the observed LF (see Liebert
et al., 1988)). A zero $BC_V$ will yield $\sim 1.5$ Gyr {\it older} ages (see Hernanz et al. 1994), for both the disk and the halo.}.
The crosses correspond to a 90\% exclusion confidence level in the
limit of detection, i.e. the possibility to see at least two WDs above this line
whereas none has been detected is rejected at the
2-$\sigma$ level (see e.g. \S IV of Liebert et al. 1988).
The solid
 lines show the {\it halo} WDLFs
for halo ages $t_h=14,16,18$ and 20 Gyr, normalized to $X_{WD}=1, 2, 4$ and 8\% respectively, for calculations done with IMF1. For the distribution of
progenitors corresponding to this IMF, differentiation at
crystallization in the WD interiors
leads to a bump in the halo WDLF in the range $-5\wig <\log (L/L_\odot)\wig < -4$, therefore ruling out substantial WD mass fractions. This shows convincingly
the importance of a {\it complete} treatment of crystallization in WD cooling.
Calculations with no carbon/oxygen differentiation will underestimate the number of WDs by more than a factor $\sim 5$,
for a given age and luminosity. Conversely they will yield halo ages $\sim 2$ Gyr younger
for a given LF. In the same vein, an incorrect Debye treatment will change significantly the shape of the WDLF. The dashed lines correspond to the same calculations when using IMF2. The normalizations correspond now to
$X_{WD}=1.7, 8, 25$ and 50\%, for the same halo ages.
Clearly, for the IMF2, a halo WD mass fraction $\wig > 30\%$, in agreement
with the MACHO results, can not be excluded, {\it provided} a halo age $\wig > 18$ Gyr.


We have compared the star counts predicted by these WDLF's with the recent
HST observations at large magnitudes (Flynn et al. 1996), for a $1/r^3$-spheroid and a $1/r^2$ halo. {\it All} WDLFs predict {\it at most} (depending on $BC_V$) $\sim 1.4$ WD in the HST field at
the limit magnitude $I=26$ for a 100\% WD halo, and thus are consistent with the HST counts. This shows that for these scarce and faint objects, large, nearby surveys put more severe constraints than deep pencil searches. 


 \section{Conclusion}

In this Letter, we have examined the possibility for
the recent MACHO events (assuming these events are genuine microlensing events)
to be due either to brown dwarfs or to white
dwarfs. This determines directly the contribution of these objects to the
missing mass in the halo of the Galaxy. Brown dwarfs are clearly excluded
as a significant halo population.
The luminosity
function of {\it halo} white dwarfs has been calculated with the most accurate white dwarf cooling theory presently available. This WDLF is confronted to {\it all} available observational constraints on halo objects.
We show that, under the two {\it necessary conditions} that i) the IMF in the
halo differs totally from the one in the disk and exhibits a strongly peaked behaviour around $m\sim 1.5-2\,\msol$, and ii) the halo is older than $\sim 18$ Gyr,
the white dwarf mass fraction in the halo can represent $\sim 25$ to 50\% of the dark matter density, in agreement with the recent MACHO results.
This would imply
an initial {\it stellar} mass fraction $> 50\%$ and thus
an essentially {\it baryonic} halo. These results are consistent with the ones
obtained from 
galactic chemical evolution (Ryu et al. 1990), though they are
in conflict with the conclusion raised by these authors that the disk must
form no later than the halo. However, as stated by these authors, alternative scenarios in the disk formation can be advocated : the left-over gas fraction might have been ejected into the intergalactic medium, as suggested by recent observations of metal-rich hot gas in the Local Group (Suto et al. 1996)\footnote{Note also that, given the low-metallicity and thus the probably less efficient mass-loss, halo WDs may have larger masses than disk WDs,
thus resulting in a smaller fraction of gas ejecta}.
The present results are also consistent with the ones obtained
by Charlot and Silk (1995), based on the expected radiation signature in
high-redshift galactic halos. These authors considered a Hubble time $< 13$ Gyr,
and solar metallicity (i.e. slowly evolving) stars. The evolution of significantly older, i.e. highly redshifted,
low-metallicity stellar populations
will certainly be consistent with these observational constraints.

Therefore,
although providing a plausible explanation for the MACHO observations and the
halo missing mass, the present
scenario relies on the necessity to invoke a {\it very peculiar}, fine-tuned IMF.
These calculations illustrate the difficulty to reconcile the recent MACHO results with
other observational constraints.
A detailed analysis of the OGLE, MACHO
and EROS results, in the context of a
consistent model for the Galaxy, will be presented in a forthcoming paper
(M\'era, Chabrier and Schaeffer 1996b).
\vfill \eject

\section*{References}

\bib Adams, F.C. and Laughlin, G., 1996, submitted to \apj

\bib Alcock C. et al., 1993, {\it Nature}, 365, 621

\bib Alcock C. et al., 1995, {\it Phys. Rev. Lett.}, 74, 2867

\bib Alcock C. et al., 1996, preprint astroph-9604176

\bib Aubourg E. et al., 1993, {\it Nature}, 365, 623



\bib Bahcall J.N., Flynn C., Gould A. and Kirhakos S., 1994, {\it ApJL}, 435,
L51

\bib Bergeron, P., Saumon, D. \& Wesemael, F., 1995, \apj, 443, 764


\bib Chabrier, G., 1993, \apj, 414, 695

\bib Charlot, S. and Silk, J., 1995, \apj, 445, 124

\bib Dahn, C.C., Liebert J., Harris H.C. \& Guetter H.H.,
1995, in {\it The bottom of the main sequence
and beyond}, Ed. C.C. Tinney, Springer

\bib Flynn C., Gould A. and Bahcall J.N., 1996, \apjl 466, 55

\bib E. Garcia-Berro, M. Hernanz, J. Isern, G. Chabrier, L. Segretain and  R. Mochkovitch, 1996, {\it A\&AS}, 117, 13

\bib Gates E.I., Gyuk G. and Turner M.S., 1995, {\it ApJL.}, 449, L123

\bib Hegyi, D.J. and Olive K.A., 1986, \apj 303, 56

\bib
M. Hernanz, E. Garcia-Berro, J. Isern, R. Mochkovitch, L. Segretain and G.
Chabrier, 1994 {\apj} 434, 652

\bib E.M. Hu, J-S Huang, G. Gilmore and L.L. Cowie, 1994, {\it Nature}, 371, 493

\bib Iben I. Jr. \& Tutukov, A.V., 1984 \apj 282, 615


\bib Larson, R.B., 1986, {\it MNRAS}, 218, 409

\bib Liebert, J., Dahn, C.C., and Monet, D.G. 1988, \apj, 332, 891

\bib M\'era, D., Chabrier, G. and Schaeffer, R., 1996a, {\it Europhysics Letters} 33, 327

\bib M\'era, D., Chabrier, G. and Schaeffer, R., 1996b, in preparation

\bib Monet, D.G. et al., 1992, \aj, 103, 638

\bib Olive, K.A., 1986, \apj 309, 210

\bib Ryu, D., Olive K.A., and Silk, J., 1990, \apj 353, 81


\bib Sahu, K., {\it Nature}, 1994, 370, 275

\bib Segretain, L. \& Chabrier, G., 1993, A\&A 271, L13

\bib L. Segretain, G. Chabrier, M. Hernanz, E. Garcia-Berro, J. Isern and
R. Mochkovitch, 1994 \apj 434, 641

\bib Suto, Y., Makishima, K., Ishisaki, Y., \& Ogasaka, Y., 1996, \apj, in press

\bib Stevenson, D.J., 1980, {\it Journal de Physique} 41, C2-61

\bib Tamanaha, C.M., Silk, J., Wood, M.A. and Winget, D.E., 1990, \apj,
358, 164

\bib vonHippel, T., Gilmore, G. \& Jones, D.H.P., 1995, {\it MNRAS} 273, L39

\bib Wood, M.A., 1992, \apj, 386, 529

 \vfill \eject


\vfill \eject \onecolumn

\centerline {\bf FIGURE CAPTIONS} \vskip1cm

\indent{\bf Figure 1 :} Halo initial mass function. The dotted line illustrates the M-dwarf MF $dM/dm\propto m^{-2.2}$ (M\'era et al. 1996a).
The solid line is the IMF $\phi(m)=A\,e^{-(\bar {m}/ m_{inf} )^{\beta_1}}
m^{-\beta_2}$ with
$\bar {m}=2.0,\beta_1=2.2$, $\beta_2$=5.15
(IMF1). The dot-dashed line
is the same IMF with
$\bar {m}=2.7,\beta_1=2.2$, $\beta_2=5.75$ (IMF2).

\vskip1cm

\indent{\bf Figure 2 :}
White dwarf luminosity function (pc$^{-3}$ M$_{bol}^{-1}$). Empty circles : Liebert et al. (1998).
Filled circles : high-velocity WDs (Liebert et al. 1988). Crosses : limit of detection at the 2-$\sigma$ level (see text). Dotted line : disk WDLF
for $t_d=10.5$ Gyr. Solid lines : halo WDLF for $t_h=14,16,18,20$ Gyr
and $X_{WD}=1,2,4,8\%$,
from left to right, with IMF1. Dashed line : halo WDLF for $t_h=14,16,18,20$ Gyr
and $X_{WD}=1.7,8,25,50\%$,
with IMF2.
 \vskip1cm

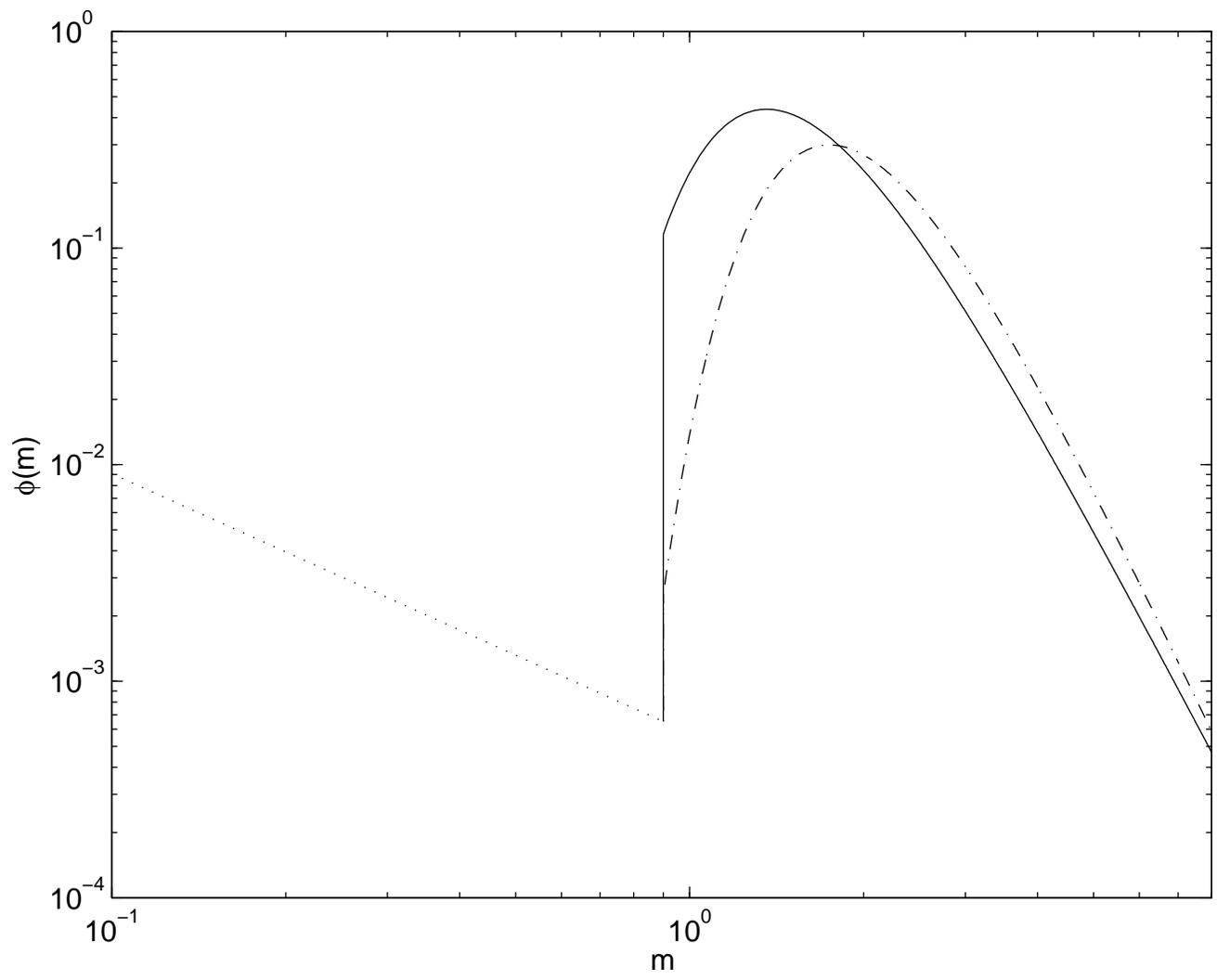
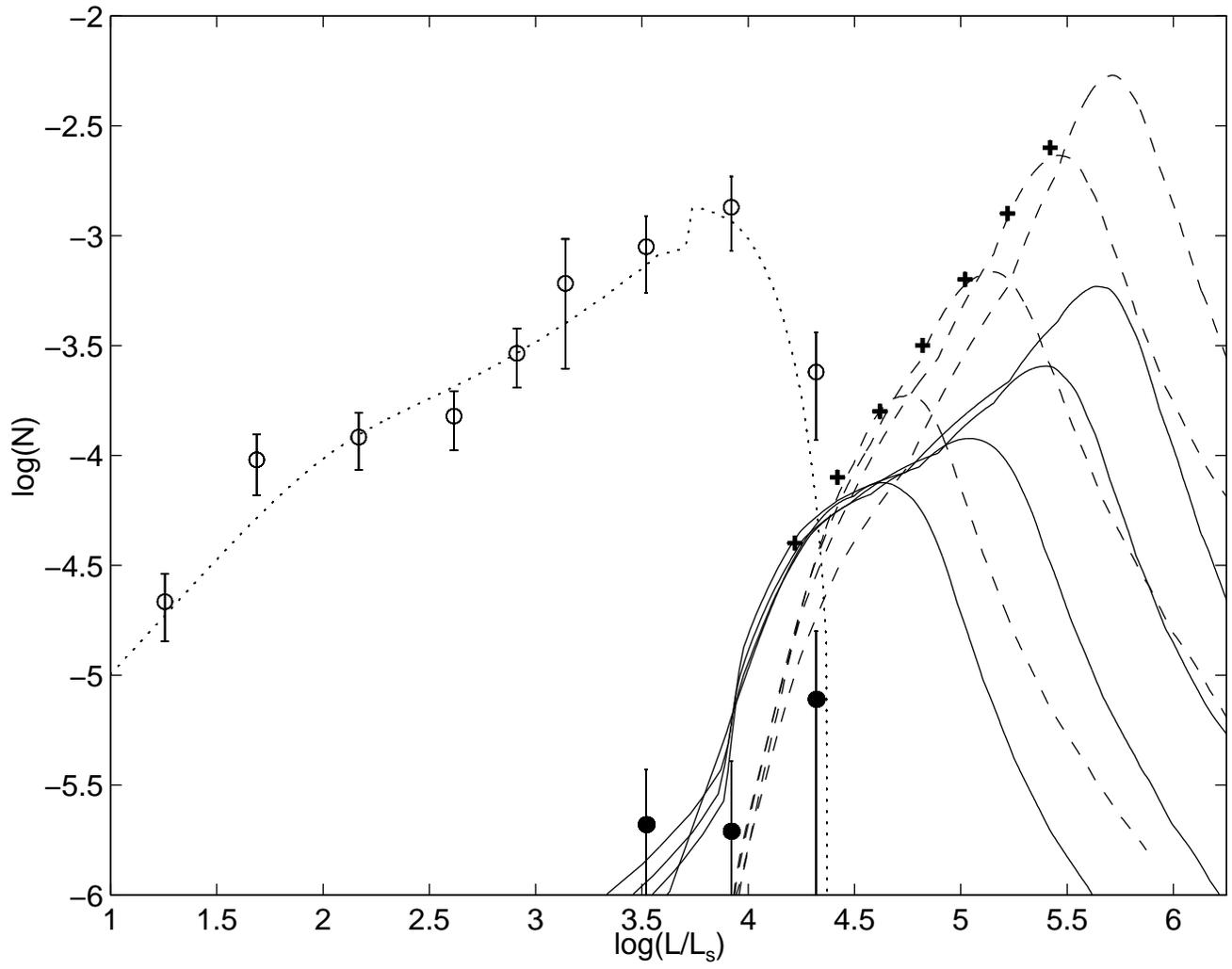
\end{document}